%% file: x2lee.tex
\documentclass{elsart2}
\journal{Physics Letters {\bf B}}
\usepackage{amssymb}
\usepackage[pdftex]{graphicx}
\usepackage{float}
\usepackage{psfrag}
\usepackage[latin1]{inputenc}

\begin{document}

%
%

\centerline{EUROPEAN ORGANIZATION FOR NUCLEAR RESEARCH}
\vspace*{1mm}
\flushright{CERN-PH-EP/2007-004\\ 21 February 2007 \\}
\vspace*{4mm}
\begin{frontmatter}
\title{First observation and branching fraction and decay parameter
measurements of the weak radiative decay
$\Xi^0\rightarrow\Lambda e^+e^-$}
\date{}
\vspace*{-5mm}
\input{authorlist_cernep}
Submitted to Physics Letters B
\vspace*{\fill}

%
%

\begin{abstract}
The weak radiative decay $\Xi^0\rightarrow\Lambda e^+e^-$
has been detected for the first time.  We find 412 candidates in the
signal region, with an estimated background of $15\pm 5$ events.  We
determine the branching fraction ${\cal{B}}(\Xi^0\rightarrow\Lambda e^+e^-) = 
[7.6\pm 0.4({\rm stat})\pm 0.4({\rm syst})\pm 0.2({\rm norm})]\times 10^{-6}$,
consistent with an internal bremsstrahlung process, and
the decay asymmetry parameter $\alpha_{\Xi\Lambda ee} = -0.8\pm 0.2$,
consistent with that of $\Xi^0\rightarrow\Lambda\gamma$.  The charge
conjugate reaction $\overline{\Xi^0}\rightarrow\overline{\Lambda} e^+e^-$
has also been observed.
\end{abstract}

\end{frontmatter}


\setcounter{footnote}{0}

%
%

\section{Introduction}

Since the discovery of hyperons, their (weak) radiative decays have
held particular interest~\cite{BE65,FE58}.
Still, the precise nature of the decays
themselves remains an open question~\cite{LA95,JE01}.

Reliable techniques to predict branching ratios remain elusive.
Furthermore, because SU(3) symmetry is broken only weakly in this regime,
weak radiative decays should
approximately conserve parity~\cite{HA64}.  Consequently, the asymmetries
of decay angular distributions should be small.  However,
results from experiments indicate
a relatively large (negative) asymmetry in every mode
investigated~\cite{YA06}.  A number of models have been proposed to
explain this apparent discrepancy~\cite{ZE00}.  Experimental results
tend to favor pole models or models based on chiral perturbation
theory, which correctly find the sign of the asymmetry.
Recently, a resolution of at least part of the puzzle has been
offered~\cite{ZE06}.

When the NA48 Collaboration undertook investigations with a
high-intensity $K^0_S$ beam
in 2002, trigger strategies for identifying radiative
hyperon decays were included from the outset.
The production over the full course of the run of more than
$3\times10^9$ neutral cascades, $\Xi^0(1315)$,
offered NA48 unmatched sensitivity for the
study of such decays.\footnote{The $\overline{\Xi^0}$
production rate was about 1/11 that of $\Xi^0$. This Letter presents
numerical results for the $\Xi^0$ only.}

This Letter details the measurement with these data
of the weak radiative hyperon decay
$\Xi^0 \rightarrow \Lambda e^+e^-$.  This is the first measurement of this
decay channel.  If one assumes an inner
bremsstrahlung-like mechanism producing the $e^+e^-$ pairs,
the expected rate for this process
may be estimated naively assuming the (virtual) photon
converts internally (Dalitz decay) or by using the machinery of
QED as carried out in rate predictions for
$\Sigma^0\rightarrow\Lambda e^+e^-$~\cite{BE65,FE58}.
The results give a range from about $1/182$ to $1/160$ of the rate of
$\Xi^0\rightarrow\Lambda\gamma$, or $(6.4-7.3)\times 10^{-6}$.
Such a process should exhibit a decay asymmetry like that in
$\Xi^0\rightarrow\Lambda\gamma$.

\section{Data}

The signal was sought among events containing one 
$\Lambda$ (which decayed in-flight to a high-momentum proton and a
much lower momentum $\pi^-$), one electron, and one positron, all in time.
An additional in-time photon was required for the normalization channel,
($\Xi^0 \rightarrow \Lambda \pi^0$, $\pi^0\rightarrow e^+e^-\gamma$),
a relatively abundant process whose final state is similar to that of the
signal and which was selected via the same trigger tree as the
signal channel.

\subsection{Beam line and detector}

The NA48 beam line
was designed to produce and transport both $K^0_L$ and $K^0_S$
beams simultaneously~\cite{BA02}.  For the 2002 run, in order to
increase dramatically the intensity of the $K^0_S$ beam, 
the $K^0_L$ target was removed and the $K^0_L$ beamline
blocked, the proton flux on the $K^0_S$ target was greatly increased,
and a 24 mm platinum absorber was placed after the Be target to
reduce the photon flux in the neutral beam.
An additional sweeping magnet was installed across the 5.2-meter long
collimator, which, tilted at 4.2 mrad relative to the incoming
proton beam, selected a beam of long-lived neutral particles ($\gamma$,
$n$, $K^0$, $\Lambda$, and $\Xi^0$).
In each 4.8 s spill, occurring every 16.2 s,
$\sim 5\times 10^{10}$ protons impinged on the target.
Approximately $2\times 10^4$
$\Xi^0$s, with momenta between 60 and 220 GeV/$c$, decayed
in the fiducial volume downstream of the
collimator each spill.

The detector for the 2002 run was identical to that used for NA48's
measurement of direct CP-violation~\cite{BA02}, except that the tagging
counter immediately after the last collimator was removed.

The neutral beam exited the final collimation into an evacuated tank,
approximately 90 m in length, terminated by a 
Kevlar window 0.3\% of a radiation length thick.
The detector was arrayed immediately downstream of this window.

A magnetic spectrometer followed the decay volume.  It consisted of four
drift chambers, two before and two after an analyzing magnet which provided
a transverse momentum kick of 265 MeV/$c$ in the horizontal plane.  The
chambers were identical, with two planes of sense wires in each of four
orientations $(x,~y,~u,~v)$, vertical, horizontal, and at $\pm 45^\circ$.
The $u$ and $v$ wires of the third chamber were the only ones left
un-instrumented.
Track-time resolution was about 1.4 ns.
Space-point resolution was approximately 150 $\mu$m in each
projection, and the momentum resolution (with $p$ in GeV/$c$) was
$\sigma_p/p = 0.48\% \oplus 0.015\%\times p$.
The resulting $m_{\pi^+\pi^-}$ resolution in $K^0_S\rightarrow\pi^+\pi^-$
decays was 3 MeV/$c^2$.

A liquid krypton calorimeter (LKr) detected and measured the
energy and position of electromagnetic showers.
Its active region was divided transversely
into approximately 2 cm $\times$ 2 cm cells, and
its depth was 27 radiation lengths.  Its single-shower time resolution was
less than 300 ps; its transverse position resolution was better
than 1.3 mm for a single photon of energy greater than 20 GeV; and its
energy resolution~\cite{UN00} was
$\sigma(E)/E = 3.2\%/\sqrt{E} \oplus 9\%/E \oplus 0.42\%$,
where $E$ is in GeV.  The resulting $m_{\gamma\gamma}$ resolution in
$\pi^0\rightarrow\gamma\gamma$ decays was approximately 1 MeV/$c^2$.

The sensitive region of the electromagnetic calorimeter primarily
constrained the fiducial volume of the experiment.  Seven rings
of scintillation counters bounded, in projection, the edges of
this acceptance region, and the last two
rings acted as trigger vetoes of extraneous activity.

A scintillator hodoscope, comprised of segmented horizontal and vertical
strips arranged in four quadrants and located between the downstream end
of the spectrometer and the upstream face of the calorimeter served as
a zeroth-level charged-track trigger.  Beyond the electromagnetic calorimeter
stood an iron-scintillator sandwich hadron calorimeter and three layers
of muon counters, each shielded by an iron wall.

The entire detector array was sampled every 25 ns.  An
event trigger initiated a readout of information within a 200 ns window
around the trigger time.
In this way, time sidebands allowed
investigations of accidental activity.

The experiment employed a multi-level trigger designed to maximize
flexibility while minimizing pile-up, dead-time losses, and the collection
of uninteresting events.  To be included in the present analysis, events
passed the lowest level hardware trigger if a horizontal-vertical
coincidence occurred in at least one quadrant of the scintillator hodoscope,
there were no in-time hits in the veto rings,
at least three views in the first drift chamber registered more than two
hits (as required in the case of more than one track),
and either the energy in the
electromagnetic calorimeter exceeded 15 GeV or the total energy in the
electromagnetic and hadron calorimeters exceeded 30 GeV.
The next level trigger required more than one
track to have passed through the spectrometer forming one or more
good vertices\footnote{A good vertex is defined, in this context, as
the occurrence of two tracks passing within 5 cm of one another between
the target and the first drift chamber.}.
The highest level trigger, an offline software
cull, passed events in which a two-track invariant mass was
consistent with that of a $\Lambda$ and contained at least
one high-energy cluster in the calorimeter not 
associated with either track forming the $\Lambda$.

A downscaled sample of minimum bias events was collected concurrently with
the physics data.  Complete trigger information was available for these
events, so trigger efficiencies could be
measured.  The relative fraction of events containing all signal final state
particles that passed the required triggers
was $\epsilon_{\rm trig}^{\rm sig} = (96.5\pm 0.2)\%$,
while the relative fraction of normalization events
was $\epsilon_{\rm trig}^{\rm norm} = (97.1\pm 0.2)\%$.

\subsection{Event selection criteria}

From events passing all trigger levels, those containing exactly four
charged tracks, two of each charge sign, that passed well within the fiducial
volumes of the first and fourth drift chambers were kept for further analysis.

Signal event simulation showed that 99\% of final state pions,
electrons, and positrons had momenta of less than 30 GeV/$c$.
A track with momentum greater than 3 GeV/$c$ and associated shower
energy within 5\% of this momentum was identified as an electron or
positron, depending on charge.  A positive track
whose momentum was greater than 30 GeV/$c$ and either had
no associated electromagnetic shower or the shower energy to momentum ratio
was less than 0.8 was identified as a proton.
If no such track was found, or if there were not both an electron and
a positron identified, the event was abandoned.
If the final track had a momentum greater than 4 GeV/$c$, but not more than
1/3.7 that of the proton track,
it was identified as a pion.  Otherwise, the event was abandoned.

The tracks associated with the proton and pion had to be separated by
at least 5 cm in the first drift chamber and their detection times
had to be within 2 ns.  If not, the event was abandoned.
The distance-of-closest-approach (doca) of the two tracks
when projected back towards the target was
required to be less than 2.2 cm, and the longitudinal position
of this doca had to lie between 4 and 40 meters down stream of the target
for the event to be further considered.
The momentum vectors of the two tracks were projected, with respect to
a reference frame centered on the beam axis,
from their positions in the first drift chamber onto the face of the LKr.
These projections were weighted by the relativistic energies of
the particles associated with the respective tracks, added vectorially, and
then normalized to the energy sum of the two particles.
The result, a quantity called
the center of gravity (COG), had to be greater than 8 cm to ensure that
a parent of the two tracks was unlikely to have been directly produced in the
target.  The COG of a directly produced particle should be small.

The invariant mass of surviving proton and pion candidate pairs
was calculated.  If the result differed from the nominal
mass of the $\Lambda$ by more than 3 MeV/$c^2$ (approximately $3\sigma$),
the event was abandoned.

The electron and positron tracks
had to have times within 2 ns and a spatial
separation in the first drift chamber of at least 2.5 cm.
The latter requirement rejects conversions in the Kevlar window.
Any unassociated shower in the calorimeter with energy
above 1.5 GeV disqualified the event as a signal candidate.

A shower of between 3 and 120 GeV
in the electromagnetic calorimeter was considered a photon
candidate for the normalization channel if it was unassociated with any
track, centered within the fiducial volume of the
detector at least 5 cm from a dead cell, and isolated from any other
shower.

Finally, for both signal and normalization channels,
the event COG, which ideally
would be 0 (see above), had to be equal to or less than 6 cm.

A signal (normalization) region was defined as
$2\sigma$ either side of the nominal $\Lambda e^+ e^-(\gamma)$
invariant mass, where $\sigma_m = 1$ MeV/$c^2$.  For the $\Lambda$,
the $p\pi$ invariant mass was used.
Selection from the entire data set
according to these criteria resulted in
412 signal candidates and 29522 normalization events reconstructed.

\section{Acceptance and reconstruction efficiency}

The product of geometrical acceptance $(A)$ and selection criteria efficiency
$(\epsilon)$ was determined with a Monte Carlo simulation.
Nearly $10^5$ signal-like events were generated
according to a two-body model of a $\Lambda$ and a virtual photon.
The model included the decay parameter $\alpha = -0.78$,
found for the decay $\Xi^0\rightarrow\Lambda\gamma$~\cite{LA04},
and a $1/m_{ee}^2$ energy distribution for the converting photon,
as would be the case for inner conversion.
In this way, the model was intended to represent inner
bremsstrahlung production.
Generated events were stepped through a GEANT simulation of the
NA48 detector and analyzed as real data,
with the result:  $(A\times\epsilon)_{\rm sig}=(2.69\pm 0.05)\%$.
For the normalization channel, about $160\times 10^6$ events (about
$7\times$ the measured flux) were generated with the latest PDG values
for the decay parameters incorporated~\cite{YA06}.  The result of
the detector simulation and reconstruction was $(A\times\epsilon)_{\rm norm}=
(0.1251\pm 0.0003)\%$.  Radiative corrections, using PHOTOS~\cite{BA94},
were included,
as was a $\Xi^0$ polarization of $-10\%$ for signal
generation.\footnote{This polarization value is consistent with that
reported by other experiments~\cite{AB06} and with indications from an ongoing
study of the NA48 beam.}

\section{Background}

Two sources of background were identified:  physics and accidentally in-time
combinations.

\subsection{Physics backgrounds}

\subsubsection{$\Xi^0\rightarrow\Lambda\pi^0$}
The $\Xi^0$ decays predominantly to $\Lambda\pi^0$.  If the $\pi^0$
Dalitz-decays, and the photon goes undetected, the final state is that
of the signal.  Similarly, if the $\pi^0$ decays via the double-Dalitz
mechanism, and an electron and a positron go undetected, the final state
is again that of the signal.  Finally, the $\pi^0\rightarrow e^+e^-$
decay results in an
irreducible background, but its rate is very small.
Simulations of each of these channels at about seven times
the flux lead to
estimates of $4.6\pm 0.8$, $0.1\pm 0.1$, and $1.2\pm 0.4$ events,
respectively, infiltrating the signal region.

\subsubsection{Kaon decays}
The flux of neutral kaons was an order of magnitude larger than that of
the $\Xi^0$.  The decay $K_S^0\rightarrow \pi^+\pi^-e^+e^-$ has a branching
fraction of $4.7\times 10^{-5}$.  If one of the pions
met the requirements of a proton in this analysis,
and the resulting $m_{p\pi} \approx m_{\Lambda}$,
then this process would mimic the signal.  Simulation with twice the
flux of such events demonstrated that an explicit mass cut
$\mid m_{\pi\pi ee} - m_{K_S^0}\mid > 0.015$ GeV/$c^2$
eliminated essentially any trace of this background with negligible impact
on signal-finding efficiency.  The decay chain $K_L^0\rightarrow
\pi^+\pi^-\pi^0$, $\pi^0\rightarrow e^+e^-\gamma$, has a product
branching ratio of about $1.5\times 10^{-3}$.  The $K_L^0$ lifetime and their
typical momentum of 80 GeV/$c$ mean that about 4\% of them
decay in the experiment's
decay volume.  For these to become a background to the $\Lambda ee$ signal,
a pion would have to be mistaken as a proton and the invariant mass of it
combined with that of the other pion would have to be close to that of
the $\Lambda$.  In addition, the photon would have to go undetected.  Because
of this last condition, an explicit kaon mass cut would be ineffective in
reducing the background.  On the other hand, the efficiency for this chain
appearing in the signal region is correspondingly reduced and the COG is
smeared out.  We estimate on the basis of Monte Carlo simulation that
$2\pm 2$ such events will populate the signal region.

\subsubsection{Accidentally in-time combinations}
\begin{figure}[t]
\begin{center}
\includegraphics[scale=0.6]{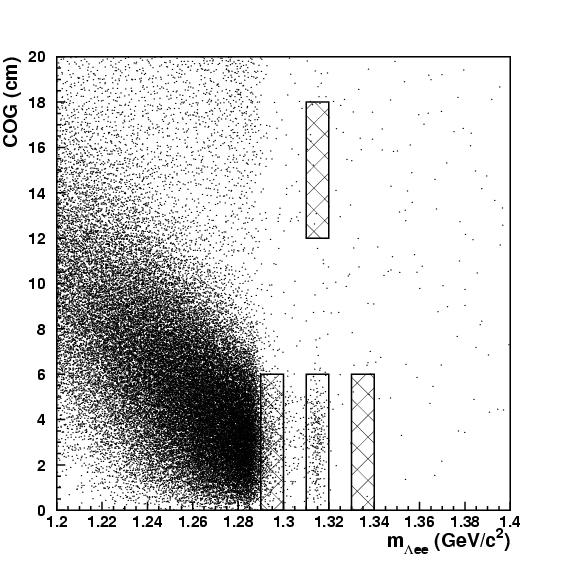}
\caption{\label{fig:cvm}COG versus $m_{\Lambda ee}$ after all other selection
criteria were imposed. The three hatched boxes are side-band regions.  The
signal region is the open box at low COG around $m_{\Xi^0}$. The side-band
regions at high mass-low COG and high COG were used to estimate
accidental and combinatoric backgrounds in the signal region.
All three side-band regions
were used in the subtraction of background under the decay-angle distribution
(see text).}
\end{center}
\end{figure}

We estimated the contamination by accidental coincidences four ways:

\begin{enumerate}
\item Running the same analysis on the data, but requiring that
the final-state leptons have the same charge.
\item  Requiring that at least one track or shower be between 10 and
20 ns out-of-time and scaling appropriately.
\item Taking events with $m_{p\pi}$ values between 7 and 10 standard
deviations from the central value $(m_\Lambda)$ and computing $m_{\Lambda ee}$.
\item Defining two ``side-band'' regions, one along each axis
in COG-versus-$m_{\Lambda ee}$ space [see, in Figure~\ref{fig:cvm},
the hatched rectangles at high COG and high mass;
each region has the same ``area'' as the
signal region, the open rectangle in the figure].
\end{enumerate}
These approaches, which are not independent, yielded between 1 and 9
events in the signal region; we take the
number to be $7\pm 5$ events.

In conclusion, combining the physics backgrounds with those attributed to
accidentals and combinatorics, the estimated number of background events
in the signal region is $15\pm 5$ [see Table
~\ref{tab:bkgd} for a summary of the background estimation].

\begin{table}[htb]
\caption{\label{tab:bkgd}Sources of expected background events.}
\begin{tabular}{|l|c|}\hline
\multicolumn{1}{|c|}{Source} & Estimate \\ \hline
$\Xi^0\rightarrow \Lambda\pi^0$, $\pi^0\rightarrow e^+e^-\gamma$ & 
$4.6\pm 0.8$ \\
$\Xi^0\rightarrow \Lambda\pi^0$, $\pi^0\rightarrow e^+e^-e^+e^-$ &
$0.1\pm 0.1$ \\
$\Xi^0\rightarrow \Lambda\pi^0$, $\pi^0\rightarrow e^+e^-$ &
$1.2\pm 0.4$ \\
Kaon Decays & $2\pm 2$ \\ 
Accidentals \& Combinatorics & $7\pm 5$ \\ \hline
{\bf TOTAL} & {\boldmath$15\pm 5$} \\ \hline
\end{tabular}
\end{table}

The background contamination of the normalization sample was estimated from
the tails of the $m_{ee\gamma}$ spectrum, which peaks sharply at
$m_{\pi^0}$.  Including a linear extrapolation under the mass peak, the number
was estimated to be $428\pm 258$.

\section{$\Xi^0$ Flux}

The total number of $\Xi^0$ produced during the run
was estimated by fully reconstructing
$\Xi^0\rightarrow \Lambda\pi^0$, $\pi^0\rightarrow e^+e^-\gamma$ events
without a longitudinal vertex position cut and
using the equation

\begin{equation}
\Phi_{\Xi^0} = \frac{N_{norm} - N_{norm~bkgd}}{(A\times\epsilon)_{\rm norm}
\epsilon_{\rm trig}^{\rm norm}
{\cal{B}}(\Xi^0\rightarrow \Lambda\pi^0){\cal{B}}(\Lambda\rightarrow p\pi^-)
{\cal{B}}(\pi^0\rightarrow e^+e^-\gamma)}
\end{equation}
From the entire data set, $29522$ such events were reconstructed.
After background subtraction, this gives an integrated flux of

\begin{displaymath}
\Phi_{\Xi^0} = (3.15\pm 0.03\pm 0.08)\times 10^9.
\end{displaymath}
The first uncertainty is due to statistics, and the second is from branching
fraction uncertainties, primarily that on
${\cal{B}}(\pi^0\rightarrow e^+e^-\gamma)$.

\begin{table}[htb]
\caption{\label{tab:fl}Quantities that entered into $\Xi^0$
flux calculations.}
\begin{tabular}{|l|r|} \hline
No. of events in signal region & 29552 \\ \hline
Estimated no. of background events & $428\pm 258$ \\ \hline
$(A\times\epsilon)_{\rm norm}$ & $(0.1251\pm 0.0003)\%$ \\ \hline
$\epsilon_{\rm trig}^{\rm norm} $ & $(97.1\pm 0.2)\%$\\ \hline
${\cal{B}}(\Xi^0\rightarrow \Lambda\pi^0)$ & $0.9952\pm 0.0003$ \\ \hline
${\cal{B}}(\Lambda\rightarrow p\pi^-)$ & $0.639\pm 0.005$ \\ \hline
${\cal{B}}(\pi^0\rightarrow e^+e^-\gamma)$ & $0.01198\pm 0.00032$ \\ \hline
\end{tabular}
\end{table}

\section{Results}

\begin{figure}[ht]
\begin{center}
\includegraphics[scale=0.6]{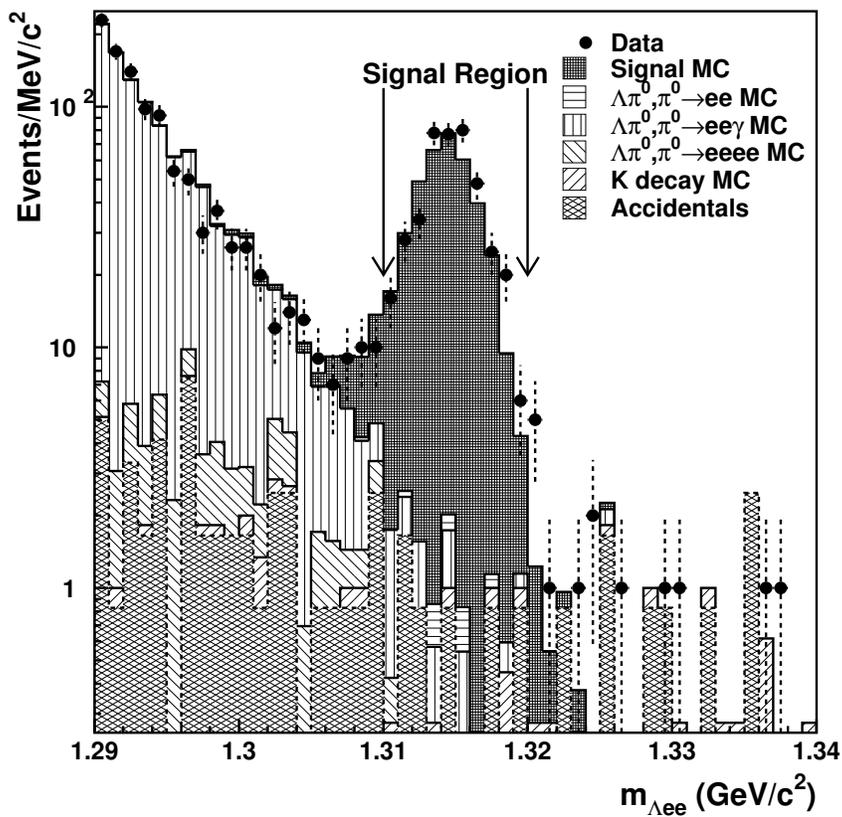}
\caption{\label{fig:mlee}$m_{\Lambda ee}$ after all selection criteria.
Arrows indicate signal region.  Stacked in various hatchings (see
legend) are the estimated sources of background.}
\end{center}
\end{figure}

At the end of the analysis, 412 events were found in the signal
region [see Figure~\ref{fig:mlee}].

\subsection{$m_{ee}$ spectrum}

\begin{figure}[t]
\begin{center}
\includegraphics[scale=0.6]{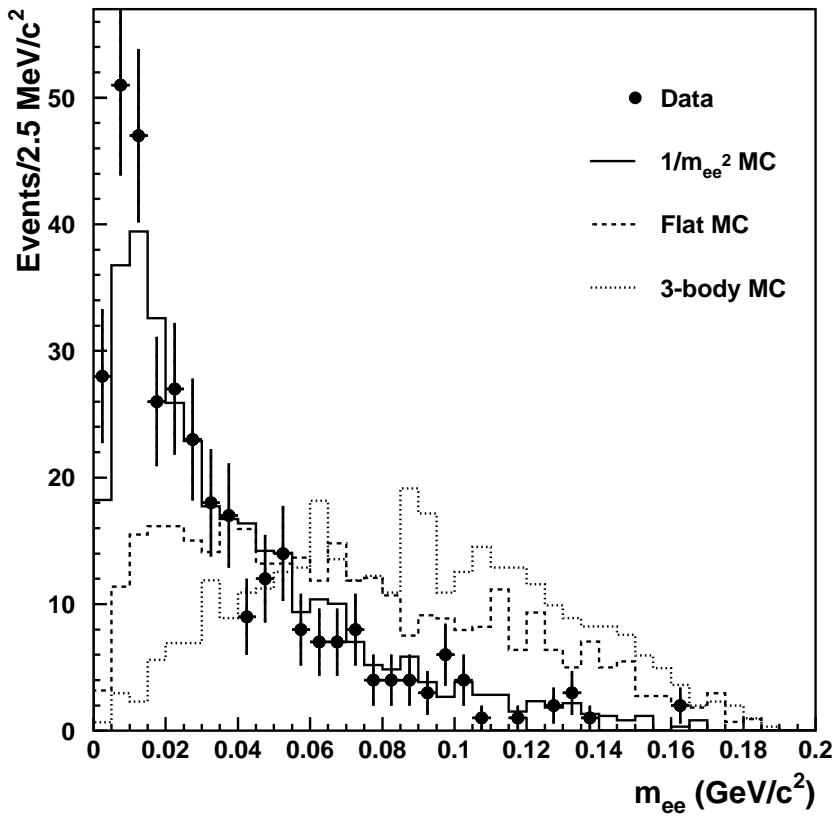}
\caption{\label{fig:mee}
Reconstructed $m_{ee}$ spectra from data (points), $1/m_{ee}^2$ (solid line),
2-body flat (dashed line), and 3-body phase space (dotted line).  The
distributions from simulated data have been normalized to
contain the same number of events in the signal region of the data,
without background subtraction.}
\end{center}
\end{figure}

The associated $m_{ee}$ distribution is consistent with a
$1/m_{ee}^2$ shape [see Figure~\ref{fig:mee}],
and we consider only this model (presumably
inner bremsstrahlung) in determining of the branching
fraction, including systematic uncertainties.

\subsection{Branching fraction}

Given the background
estimate, efficiencies, and flux discussed above, and the PDG
$\Lambda\rightarrow p\pi^-$ branching ratio [see Table~\ref{tab:br}],
the branching ratio for $\Xi^0\rightarrow\Lambda e^+e^-$ is determined to be

\begin{displaymath}
{\cal{B}}(\Xi^0\rightarrow\Lambda e^+e^-) = (7.6\pm 0.4) \times 10^{-6}
\end{displaymath}
where the uncertainty here is statistical only.

\begin{table}[htb]
\caption{\label{tab:br}Quantities that entered into branching
fraction calculations.}
\begin{tabular}{|l|r|} \hline
No. of events in signal region & 412 \\ \hline
Estimated no. of background events & $15\pm 5$ \\ \hline
$(A\times\epsilon)_{\rm sig}$ & $(2.69\pm 0.05)\%$ \\ \hline
$\epsilon_{\rm trig}^{\rm sig} $ & $(96.5\pm 0.2)\%$\\ \hline
$\Xi^0$ flux & $(3.14\pm 0.03)\times 10^9$ \\ \hline
${\cal{B}}(\Lambda\rightarrow p\pi^-)$ & $0.639\pm 0.005$ \\ \hline
\end{tabular}
\end{table}

A check for $\overline{\Xi^0}\rightarrow\overline{\Lambda}e^+e^-$
found a clear peak of 24 events, of which, roughly, as many as 7 may be
background, in the invariant mass plot.  This number is
consistent with the $\Xi^0$ branching fraction and relative
$\Xi^0$ and $\overline{\Xi^0}$ production rates.  The
production mechanism, kinematics and backgrounds of the
$\overline{\Xi^0}$, however, differ from those of the $\Xi^0$,
and no further consideration of this charge-conjugate channel is given here.

\subsection{Systematic uncertainties}

Analysis selection criteria were varied when looking at the data and
when determining reconstruction efficiencies.  The branching fraction
result was most sensitive to the treatment of the reconstructed $\Xi^0$
vertex and backgrounds from the $\Xi^0\rightarrow \Lambda\pi^0$ channel
in relation to $m_{ee}$.  No cut was
placed on the longitudinal position of the $\Xi^0$ vertex.  Requirements
varying the minimum longitudinal position of the vertex in 6-m intervals
beginning before the target (to account for resolution effects)
resulted in branching fraction changes of between 0.2\% and 3\%.  We
assign the highest variation $(\pm 3\%)$ as a systematic error.

It was possible to eliminate nearly all physics backgrounds by excluding
signal events with $0.100~{\rm GeV}/c^2 < m_{ee} < 0.135~{\rm GeV}/c^2$,
which, according to signal Monte Carlo, reduces the reconstruction
efficiency by 5\%.  Cutting this region from the final data sample, and
recalculating the branching ratio, results in a shift of 1.8\%,
which was included symmetrically as a systematic uncertainty.
These, along with smaller variations in the branching fraction
resulting from other modifications of the selection
criteria, were added in quadrature to give
a systematic uncertainty of $\pm 3.6\%$ on the branching fraction.

We conservatively assign a relative $\pm 1\%$ uncertainty on the
determination of the background to account for correlations in methods for
estimating accidentally in-time events.

The branching fraction differed by about 1\% when signal and normalization
modes were simulated with and without radiative corrections,
and we include
this difference symmetrically as a systematic uncertainty.

For the $A\times\epsilon$ determinations, the $\Xi^0$ polarization of simulated
events was
set to $-10\%$.
Samples of simulated data, generated with the polarization varied
between 0\% and $-20$\% $(\pm 10\%)$, were used to recalculate the branching
fraction vary.  The largest variation among these trials was 2.7\%,
and this variation is taken symmetrically as a systematic uncertainty.

The decay asymmetry used in generating simulated signal events was
that of the process $\Xi^0\rightarrow\Lambda\gamma$~\cite{LA04}.  Our
measurement, discussed below, is in agreement with this value, but with a 25\%
uncertainty.  Varying our simulation within this 25\% range changed
the branching fraction by at most 2.5\%, and this is 
symmetrically assigned to systematic uncertainty.

The determination of the trigger efficiency and $\Xi^0$ flux were discussed
above.  The difference between trigger efficiencies for signal and
normalization channels is taken as an uncertainty, affecting the branching
ratio by 0.6\%.  An alternative, less direct,
calculation of the flux was statistically consistent
with the one described above.  The two differed by 1.9\%,
and we conservatively include, symmetrically, this amount as a
systematic uncertainty.

The total systematic uncertainty on the branching fraction,
recounted in Table~\ref{tab:su}, is
$\pm 5.7\%$, the sum in quadrature of each of the sources described.
This gives a final branching fraction of:

\begin{displaymath}
{\cal{B}}(\Xi^0\rightarrow\Lambda e^+e^-) =
[7.6\pm 0.4({\rm stat})\pm 0.4({\rm syst})\pm 0.2({\rm norm})]\times 10^{-6}.
\end{displaymath}

\begin{table}[htb]
\caption{\label{tab:su}Sources of systematic uncertainty on the
branching fraction.}
\begin{tabular}{|l|c|}\hline
\multicolumn{1}{|c|}{Source} & Fractional \\
& Uncertainty \\ \hline
Detector Acceptance & 3.6\% \\
Background & 1.0\% \\
Radiative Corrections & 1.0\% \\
Polarization & 2.7\% \\
Signal Modeling & 2.5\% \\
Trigger Efficiency & 0.6\% \\
$\Xi^0$ Flux & 1.9\% \\ \hline
{\bf TOTAL} & {\bf 5.7\%} \\ \hline
\end{tabular}
\end{table}

\subsection{Asymmetry parameter}

The angular distribution of the proton relative to the $\Xi^0$ line of
flight in the $\Lambda$ rest frame is given by~\cite{YA06}:

\begin{equation}
\frac{dN}{d\cos{\theta_{p\Xi}}} = \frac{N}{2}(1 -
\alpha_{\Xi\Lambda ee}\alpha_{\_}\cos{\theta_{p\Xi}}).
\end{equation}

\begin{figure}[t]
\begin{center}
\includegraphics[scale=0.6]{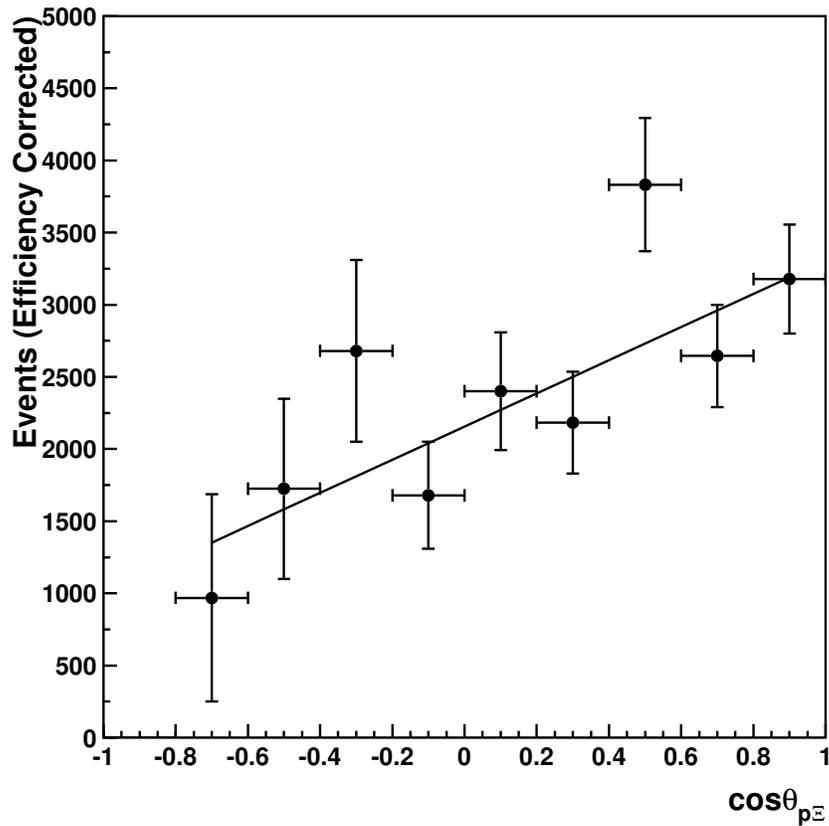}
\caption{\label{fig:cos}
Background-subtracted and acceptance-corrected
$\cos{\theta}_{p\Xi}$ distribution.  The line is the fit result.}
\end{center}
\end{figure}

The $\cos{\theta_{p\Xi}}$ spectrum from signal events
was corrected by subtracting scaled
backgrounds from the side-band regions indicated in Figure~\ref{fig:cvm} and
by dividing, bin-by-bin, the acceptance as determined from a $\Xi^0
\rightarrow \Lambda e^+e^-$ simulation where the spectrum was generated
to be flat in $\cos{\theta_{p\Xi}}$.
A two-parameter fit to this corrected spectrum gives the
product of asymmetry parameters $\alpha_{\Xi\Lambda ee}\alpha_{\_}$, where
$\alpha_{\_}$ is the asymmetry parameter for the decay $\Lambda\rightarrow
p\pi^-$.  This latter was taken to be $\alpha_{\_} = 0.642\pm
0.013$~\cite{YA06}.
The fit (over the interval $-0.8< \cos{\theta_{p\Xi}}< 1$)
[see Figure~\ref{fig:cos}]
to the data yields,

\begin{displaymath}
\alpha_{\Xi\Lambda ee} = -0.8\pm 0.2
\end{displaymath}

This is consistent with the latest published value of
$\alpha_{\Xi\Lambda\gamma} = 
-0.78\pm 0.18({\rm stat})\pm 0.06({\rm syst})$~\cite{LA04}.

\section{Summary and conclusions}

The weak radiative decay channel $\Xi^0\rightarrow \Lambda e^+ e^-$
has been identified.
Its branching fraction has been determined to be

\begin{displaymath}
{\cal{B}}(\Xi^0\rightarrow\Lambda e^+e^-) =
[7.6\pm 0.4({\rm stat})\pm 0.4({\rm syst})\pm 0.2({\rm norm})]\times 10^{-6},
\end{displaymath}
consistent with an inner bremsstrahlung-like production mechanism for the
$e^+e^-$ pair.  The consistency is further supported by the
$m_{ee}$ spectrum.  The decay parameter

\begin{displaymath}
\alpha_{\Xi\Lambda ee} = -0.8\pm 0.2,
\end{displaymath}
is consistent with that measured for $\Xi^0\rightarrow \Lambda\gamma$.

The existence of the charge conjugate reaction
$\overline{\Xi^0}\rightarrow\overline{\Lambda} e^+e^-$, has been
confirmed.

\paragraph*{Acknowledgments}
\noindent
\\[7pt]
It is a pleasure to thank technical personnel from participating laboratories
and universities, and affiliated computer centers, for their
essential contributions to constructing the apparatus, running the experiment,
and processing the data.

\end{document}

%% file: authorlist_cernep.tex
\author{NA48 collaboration}
\address{\ }
\vspace*{-7mm}
%
%
\author{J.R.~Batley},
\author{G.E.~Kalmus\thanksref{threfRAL}},
\author{C.~Lazzeroni},
\author{D.J.~Munday},
\author{M.~Patel},
\author{M.W.~Slater},
\author{S.A.~Wotton}
\address{Cavendish Laboratory, University of Cambridge, Cambridge, CB3 0HE, U.K.\thanksref{thref3}}
\thanks[threfRAL]{Present address: Rutherford Appleton Laboratory, Chilton, Didcot, Oxon, OX11~0QX, U.K.}
\thanks[thref3]{Funded by the U.K.\ Particle Physics and Astronomy Research Council.}
\author{R.~Arcidiacono\thanksref{threfMIT}},
\author{G.~Bocquet},
\author{A.~Ceccucci},
\author{D.~Cundy\thanksref{threfZX}},
\author{N.~Doble\thanksref{threfPisa}},
\author{V.~Falaleev},
\author{L.~Gatignon},
\author{A.~Gonidec},
\author{P.~Grafstr\"om},
\author{W.~Kubischta},
\author{I.~Mikulec\thanksref{threfXY}},
\author{A.~Norton},
\author{B.~Panzer-Steindel},
\author{P.~Rubin\thanksref{threfPhil}\corauthref{cor}},
\corauth[cor]{Corresponding author. {\em Email address:} prubin@gmu.edu}
\author{H.~Wahl\thanksref{threfHW}}
\address{CERN, CH-1211 Gen\`eve 23, Switzerland} 
\thanks[threfMIT]{Present address: Massachusetts Institute of Technology, Cambridge, MA~02139-4307, U.S.A.}
\thanks[threfZX]{Present address: Istituto di Cosmogeofisica del CNR di Torino, I-10133~Torino, Italy}
\thanks[threfPisa]{Present address: Dipartimento di Fisica, Scuola Normale Superiore e Sezione dell'INFN di Pisa, I-56100~Pisa, Italy}
\thanks[threfXY]{On leave from \"Osterreichische Akademie der Wissenschaften, Institut  f\"ur Hochenergiephysik,  A-1050 Wien, Austria}
\thanks[threfPhil]{On leave from University of Richmond, Richmond, VA, 23173, USA; 
                   supported in part by the US NSF under award \#0140230.
                   Present address: Department of Physics and Astronomy, George Mason University, Fairfax, VA 22030A, USA.}
\thanks[threfHW]{Present address: Dipartimento di Fisica dell'Universit\`a e Sezione dell'INFN di Ferrara, I-44100~Ferrara, Italy}
%
%
\author{E.~Goudzovski\thanksref{threfPisa}},
\author{P.~Hristov\thanksref{threfCERN}},
\author{V.~Kekelidze},
\author{L.~Litov},
\author{D.~Madigozhin},
\author{N.~Molokanova},
\author{Yu.~Potrebenikov},
\author{S.~Stoynev},
\author{A.~Zinchenko}
\address{Joint Institute for Nuclear Research, Dubna, Russian    Federation}  
\thanks[threfCERN]{Present address: CERN, CH-1211 Gen\`eve~23, Switzerland}
%
%
\author{E.~Monnier\thanksref{threfMonnier}},
\author{E.~Swallow},
\author{R.~Winston\thanksref{threfWinston}},
\address{The Enrico Fermi Institute, The University of Chicago, Chicago, IL 60126, U.S.A.}
\thanks[threfMonnier]{Present address: Centre de Physique des Particules de Marseille, IN2P3-CNRS, Universit\'e de la M\'editerran\'ee, Marseille, France.}
\thanks[threfWinston]{Also at University of California, Merced, U.S.A.}
%
%
\author{R.~Sacco\thanksref{threfSacco}},
\author{A.~Walker}
\address{Department of Physics and Astronomy, University of Edinburgh, JCMB King's Buildings, Mayfield Road, Edinburgh, EH9~3JZ, U.K.}
\thanks[threfSacco]{Present address: Department of Physics, Queen Mary University, London, E1~4NS, U.K.}
%
%
\author{W.~Baldini},
\author{P.~Dalpiaz},
\author{P.L.~Frabetti\thanksref{threfFrabetti}},
\author{A.~Gianoli},
\author{M.~Martini},
\author{F.~Petrucci},
\author{M.~Savri\'e},
\author{M.~Scarpa}
\address{Dipartimento di Fisica dell'Universit\`a e Sezione    dell'INFN di Ferrara, I-44100 Ferrara, Italy}
\thanks[threfFrabetti]{Present address: Joint Institute for Nuclear Research, Dubna, 141980, Russian Federation}
%
\newpage
\author{A.~Bizzeti\thanksref{threfXX}},
\author{M.~Calvetti},
\author{G.~Collazuol\thanksref{threfPisa}},
\author{E.~Iacopini},
\author{M.~Lenti},
\author{G.~Ruggiero\thanksref{threfCERN}}
\author{M.~Veltri\thanksref{thref7}}
\address{Dipartimento di Fisica dell'Universit\`a e Sezione dell'INFN di Firenze, I-50125~Firenze, Italy}
\thanks[threfXX]{Dipartimento di Fisica dell'Universit\`a di Modena e Reggio Emilia, I-41100~Modena, Italy}
\thanks[thref7]{Istituto di Fisica dell'Universit\`a di Urbino, I-61029~Urbino, Italy}
%
%
\author{M.~Behler},
\author{K.~Eppard},
\author{M.~Eppard\thanksref{threfCERN}},
\author{A.~Hirstius\thanksref{threfCERN}},
\author{K.~Kleinknecht},
\author{U.~Koch},
\author{P.~Marouelli},
\author{L.~Masetti\thanksref{threfBonn}},
\author{U.~Moosbrugger},
\author{C.~Morales Morales},
\author{A.~Peters\thanksref{threfCERN}},
\author{R.~Wanke},
\author{A.~Winhart}
\address{Institut f\"ur Physik, Universit\"at Mainz, D-55099~Mainz, Germany\thanksref{thref6}}
\thanks[threfBonn]{Present address: Physikalisches Institut, Universit\"at 
Bonn, 53113 Bonn, Germany}
\thanks[thref6]{Funded by the German Federal Minister for Research and Technology (BMBF) under contract 7MZ18P(4)-TP2}
\author{A.~Dabrowski},
\author{T.~Fonseca Martin\thanksref{threfCERN}},
\author{M.~Velasco}
\address{Department of Physics and Astronomy, Northwestern University, Evanston, IL 60208-3112, U.S.A.}
%
%
%
%
%
\author{G.~Anzivino},
\author{P.~Cenci},
\author{E.~Imbergamo},
\author{G.~Lamanna\thanksref{threfPisa}},
\author{P.~Lubrano},
\author{A.~Michetti},
\author{A.~Nappi},
\author{M.~Pepe},
\author{M.C.~Petrucci}
\author{M.~Piccini\thanksref{threfCERN}}
\author{M.~Valdata}
\address{Dipartimento di Fisica dell'Universit\`a e Sezione dell'INFN di Perugia, I-06100~Perugia, Italy}
%
%
\author{C.~Cerri},
\author{F.~Costantini},
\author{R.~Fantechi},
\author{L.~Fiorini\thanksref{threfLuca}},
\author{S.~Giudici},
\author{I.~Mannelli},
\author{G.~Pierazzini},
\author{M.~Sozzi}
\address{Dipartimento di Fisica, Scuola Normale Superiore e Sezione dell'INFN di Pisa, I-56100 Pisa, Italy} 
\thanks[threfLuca]{Present address: Cavendish Laboratory, University of Cambridge, Cambridge, CB3 0HE, U.K.}
%
%
\author{C.~Cheshkov},
\author{J.B.~Cheze},
\author{M.~De Beer},
\author{P.~Debu},
\author{G.~Gouge},
\author{G.~Marel},
\author{E.~Mazzucato},
\author{B.~Peyaud},
\author{B.~Vallage}
\address{DSM/DAPNIA - CEA Saclay, F-91191~Gif-sur-Yvette, France} 
%
%
%
\author{M.~Holder},
\author{A.~Maier\thanksref{threfCERN}},
\author{M.~Ziolkowski }
\address{Fachbereich Physik, Universit\"at Siegen, D-57068 Siegen, Germany\thanksref{thref8}}
\thanks[thref8]{Funded by the German Federal Minister for Research and Technology (BMBF) under contract 056SI74}
\newpage
\author{C.~Biino},
\author{N.~Cartiglia},
\author{M.~Clemencic},
\author{S.~Goy Lopez},
\author{F.~Marchetto}, 
\author{E.~Menichetti},
\author{N.~Pastrone}
\address{Dipartimento di Fisica Sperimentale dell'Universit\`a e Sezione dell'INFN di Torino, I-10125~Torino, Italy} 
%
%
%
\author{W.~Wislicki}
\address{Soltan Institute for Nuclear Studies, Laboratory for High Energy Physics, PL-00-681~Warsaw, Poland\thanksref{thref9}}
\thanks[thref9]{Supported by the Committee for Scientific Research grants 5P03B10120, SPUB-M/CERN/P03/DZ210/2000 and SPB/CERN/P03/DZ146/2002}
\author{H.~Dibon},
\author{M.~Jeitler},
\author{M.~Markytan},
\author{G.~Neuhofer},
\author{L.~Widhalm}
\address{\"Osterreichische Akademie der Wissenschaften, Institut  f\"ur Hochenergiephysik, A-1050~Wien, Austria\thanksref{thref10}}
\thanks[thref10]{Funded by the Austrian Ministry for Traffic and Research under the contract GZ 616.360/2-IV GZ 616.363/2-VIII, and by the Fonds f\"ur Wissenschaft und Forschung FWF Nr.~P08929-PHY}

%% file: x2lee.bbl
\begin{thebibliography}{MM99z}

\bibitem{BE65} J. Bernstein, G. Feinberg, and T. D. Lee, Phys. Rev.
{\bf 139}, 1650 (1965).

\bibitem{FE58} G. Feldman and T. Fulton, Nucl. Phys. {\bf 8}, 106
(1958).

\bibitem{LA95} J. Lach, P. $\dot{\rm Z}$enczykowski,
Int. J. Mod. Phys. A{\bf 10}, 3817 (1995).

\bibitem{JE01} D. A. Jensen, Nucl. Phys. B (Proc. Suppl.) {\bf 93},
22 (2001).

\bibitem{HA64} Y. Hara, Phys. Rev. Lett. {\bf 12}, 378 (1964).

\bibitem{YA06} W.-M. Yao et al. (Particle Data Group),
J. Phys. G {\bf 33}, 1 (2006) (URL: http://pdg.lbl.gov)

\bibitem{ZE00} P. $\dot{\rm Z}$enczykowski, Phys. Rev. D {\bf 62},
014030 (2000), and references therein.

\bibitem{ZE06} P. $\dot{\rm Z}$enczykowski, hep-ph/0610191.

\bibitem{BA02} J. R. Batley, et al., Phys. Lett. B {\bf 544}, 97 (2002).

\bibitem{UN00} NA48 Collaboration, G. Unal, in:  IX International
Conference on Calorimetry, October 2000, Annecy, France, hep-ex/0012011.

\bibitem{LA04} A. Lai, et al., Phys. Lett. B {\bf 584}, 251 (2004).

\bibitem{BA94} E. Barberio and Z. Was, Comput. Phys. Commun. {\bf 79}, 291
(1994).

\bibitem{AB06} E. Abouzaid, et al., hep-ex/0608007.













\end{thebibliography}
